# Origins of minimized lattice thermal conductivity and enhanced thermoelectric performance in WS$_2$/WSe$_2$ lateral superlattice

*Yonglan Hu$^{\#a}$, Tie Yang$^{\#b}$, Dengfeng Li$^a$, Guangqian Ding$^{*a}$, Chaochao Dun$^{*c}$, Dandan Wu$^d$ and Xiaotian Wang$^{*b}$*

$^a$School of Science, Chongqing University of Posts and Telecommunications, Chongqing, 400065, China.

$^b$School of Physical Science and Technology, Southwest University, Chongqing, 400715, China.

$^c$Department of Aerospace and Mechanical Engineering, University of Notre Dame, Notre Dame, Indiana 46556, USA.

$^d$Institutes of Physical Science and Information Technology, Anhui University, Hefei, Anhui 230601, China.

$^{\#}$These authors contribute equally to this work.

**ABSTRACT:** We report a configuration strategy for improving the thermoelectric (TE) performance of two-dimensional (2D) transition metal dichalcogenide (TMDC) WS$_2$ based on the experimentally prepared WS$_2$/WSe$_2$ lateral superlattice (LS) crystal. On the basis of density function theory combined with Boltzmann transport equation, we show that the TE figure of merit $zT$ of monolayer WS$_2$ is remarkably enhanced when forming into a WS$_2$/WSe$_2$ LS crystal. This is primarily ascribed to the almost halved lattice thermal conductivity due to the enhanced anharmonic processes. Electronic transport properties parallel ($xx$) and perpendicular ($yy$) to the superlattice period are highly symmetric for both $p$- and $n$-doped LS owing to the nearly isotropic lifetime of charger carriers. The spin-orbital effect causes a significant split of conduction band and leads to three-fold degenerate sub-bands and high density of states (DOS), which offers opportunity to obtain the high $n$-type Seebeck coefficient ($S$). Interestingly, the separated degenerate sub-bands and upper conduction band in monolayer WS$_2$ form a remarkable stairlike DOS, yielding a higher $S$. The hole carriers with much higher mobility than electrons reveal the high $p$-type power factor and the potential to be good $p$-type TE materials with optimal $zT$ exceeds 1 at 400K in WS$_2$/WSe$_2$ LS.

## INTRODUCTION

Thermoelectric cooling or power generation devices can realize direct conversion between heat and electricity. The efficiency of the TE conversion primarily depends on the figure of merit $zT$ of a material, $zT=S^2\sigma T/(\kappa_e+\kappa_l)$, where $S$, $\sigma$ and $T$ are the Seebeck coefficient, electrical conductivity and absolute temperature, respectively, $\kappa_e$ and $\kappa_l$ are the electronic and lattice contribution to thermal conductivity.[1-3] In the past decades, intense effort taken to improve the $zT$ of exiting complex bulk materials seems unsatisfactory since these TE coefficients are strongly coupled, e.g., the $S$ and $\sigma$ are inversely related while $\sigma$ and $\kappa_e$ are proportionally related.[4-6] Achieving high $zT$ requires high $S$, high $\sigma$ and low thermal conductivity. Thus, new breakthrough in $zT$ lies in how to control these coefficients individually. It has been suggested that low-dimensional DOS could have potential in increasing $S$ without reducing $\sigma$, meanwhile the introduced boundary or interface can scatter phonons and reduce $\kappa_l$. These concepts are quite old and the demonstration was presented in the seminal papers by Hicks and Heremans.[7-9] Nowadays, exploring competitive $zT$ in low-dimensional nano-crystals becomes a promising scenario in TE community. Potential candidates such as 2D TMDCs,[10] phosphorene,[11] silicene,[12] bismuth oxyselenide[13] have been widely investigated.

2D TMDCs draw growing interest in TE due to their superior stability and high carrier mobility. It has been found that the TE performance of 1T-type TMDCs ($M$S(Se)$_2$, $M$=Ti, Zr) are much higher than the 2H-type TMDCs ($M$S(Se)$_2$, $M$=Mo, W). For example, the optimal $zT$ of monolayer ZrS$_2$ at 300K is about 1.7,[14] while it is less than 0.1 in monolayer MoS$_2$.[15] Such a great difference in TE performance arises from the distinct phonon transport behaviors, i.e., the $\kappa_l$ at 300K for monolayer ZrS$_2$ and MoS$_2$ are 3.29 W/mK and 83 W/mK,[14,15] respectively. Based on Gu and Yang,[16] the high $\kappa_l$ of 2H TMDCs monolayers comes from two crucial factors: (i) The strong bonding stiffness leads to large span of phonon frequency and therefore the high phonon velocity, and (ii) the large atomic weight difference causes a wide frequency gap which forbids the anharmonic phonon scattering processes. Similar results was also found in monolayer WSe$_2$ and MoSe$_2$ according to the investigation of Kumar and Schwingenschlogl.[17] As the controlling synthesis of 2H TMDCs monolayers is now a mature technology, improving their TE performance are important for their potential application in TE devices, which in turn lies in how to break their lower limit of $\kappa_l$ by introducing additional impurities or interfaces.

Interestingly, Duan et al.[18,19] reported the lateral epitaxial growth of 2D heterostructures and superlattices based on 2D TMDCs, and the WS$_2$/WSe$_2$ and MoS$_2$/MoSe$_2$ LSs were well prepared. Besides, a variety of combination of LS such as MoS$_2$-WS$_2$,[20] WSe$_2$-MoS$_2$[21] were also experimentally prepared, and it is possible to turn their morphological or electronic properties. Li et al.[22] even theoretically considered the combination of Janus monolayer with LS crystal (MoSSe and WSSe). On the basis of the theory of low-dimensional TE transport and this experimental realization, it is desirable to improving the TE performance of 2H TMDCs monolayers by forming such LS crystal. As demonstrated by Hicks,[7] the quantum-well superlattices can alter $zT$ since carriers are confined to move in well while the interfaces can scatter the phonons. We have shown that the $\kappa_l$ in MoS$_2$/MoSe$_2$ LS is greatly minimized due to enhanced phonon scattering rates, as compared to the pristine monolayers.[23] Liu et al.[24] also reported the high TE performance in graphene/boron-nitride LS ascribing to the same mechanism. To date, exploring the TE transport in LS crystals based on exiting 2D monolayers is lacking, and mechanisms behind are still not fully understood. Although we have preliminary investigated the phonon transport in $MoS_2/MoSe_2$ LS, the mechanisms of electronic transport and fully TE performance of LS are still unclear. Here, we report the TE transport performance of monolayer $WS_2$ and $WS_2/WSe_2$ LS. We find the markedly inhibited phonon transport and almost remained electronic transport in $WS_2/WSe_2$ LS with the enhanced *zT* exceeds 1 at 400K in *p*-doped LS.

**COMPUTATIONAL DETAILS**

To obtain the stable structure with fully relaxed lattice parameters and atomic positions, we carry out first-principles calculations within the framework of density functional theory (DFT) using the projector-augmented-wave (PAW)[25] formalism and the generalized gradient approximation (GGA) Perdew–Burke–Ernzerhof (PBE)[26] exchange-correlation functional, as implemented in VASP.[27] The plane-wave cutoff energy is 400 eV and the Monkhorst-Pack *k*-points are 9×15×1 and 7×21×1 for monolayer $WS_2$ and $WS_2/WSe_2$ LS, respectively. The convergence criteria of the self-consistent loop is $10^{-6}$ eV. The lattices and positions are fully relaxed until the maximum force becomes less than 0.01 eV/Å. We determine that the unit cell parameters *a*(*b*) for monolayer $WS_2$ and $WS_2/WSe_2$ LS are 5.48(3.16)Å and 11.24(3.25)Å, respectively. When calculating the electronic band structure, the Monkhorst-Pack *k*-points are increased to 19×31×1 and 17×57×1 to guarantee a converged TE coefficients. Besides, the spin-orbital coupling (SOC) and modified Becke-Johnson (MBJ)[28] functional are also considered to yield the accurate effective mass and band-gap.

The electronic transport coefficients are calculated using the Boltzmann transport equation (BTE) in a constant relaxation time approximation (CRTA) as implemented in BoltzTraP.[29] A rigid band approximation[30] is used to treat doping, the Fermi level shifts up for *n*-type doping while down for *p*-type. The *S*, $\sigma$, and $\kappa_e$ are obtained by solutions of BTE in term of the transport distribution function: $\sum(\varepsilon) = \sum_{\bar{k}} v_{\bar{k}}^2 \tau_{\bar{k}} \delta(\varepsilon - \varepsilon_{\bar{k}})$,[31] where $v_{\bar{k}}$ is the group velocity of carriers with wave vector $\bar{k}$, $\tau_{\bar{k}}$ is the carrier's lifetime, and $\varepsilon_{\bar{k}}$ is the dispersion relation for the carriers.

The $\kappa_l$ is calculated by solving phonon BTE as implemented in ShengBTE.[32-34] A linearized pBTE can be defined when considering two- and three-phonon processes as the scattering sources, and also the scattering from isotopic disorder is also included in this method. Harmonic and anharmonic interatomic force constants (IFCs) are quite necessary inputs for pBTE, which are obtained from DFT calculations using converged 3×3×1 and 2×5×1 supercells for monolayer $WS_2$ and $WS_2/WSe_2$ LS, respectively. A converged cut-off distance 0.6 nm for interactive distance is used in calculating anharmonic IFCs. The harmonic IFCs are obtained by Phonopy code[35] For obtaining anharmonic IFCs and for solving pBTE, we employ the ShengBTE code,[32] based on adaptive smearing approach to the conservation of energy and on an iterative solution method. It is worthwhile to note that the four-phonon scattering is not considered in present calculation since it usually plays significant role at high-temperature and also for ultrahigh-thermal conductivity materials, according to Feng *et al*.[36] Besides, since the volume is not a well-defined quantity in 2D crystal, a thickness of *h*=6.2155 Å is applied, which is equal to the layer separation of bulk $WS_2$.[6]

## RESULTS AND DISCUSSION

Monolayer $WS_2$ and $WSe_2$ belong to the 2D hexagonal 2H-$MoS_2$ type TMDCs, as shown in Figure 1a, the unit cell comprises a S(Se)-W-S(Se) sandwich atomic sequence and the two S(Se) atoms are symmetrical to the hexagonal W-W plane, and they were prepared by Lan and Huang,[37, 38] respectively. Different from graphene, both $WS_2$ or $WSe_2$ monolayers are semiconductor with direct band-gap about 1.55eV[39] and 1.56eV [17], respectively, hence, they draw growing interest in electronic and optical devices. They are also highlighted as potential TE materials due to their superior stability and high carrier mobility, while the high $\kappa_l$ in the crystals overshadows these advantages with the optimal $zT$ less than 0.1 at room temperature.[17, 39] We set out to turn the TE transport properties by a model of LS structure. In order to construct a $WS_2$/$WSe_2$ LS, we first rebuild the hexagonal crystal of monolayer $WS_2$ in a orthorhombic crystal, as shown in Figure 1b, which allows us to explore the TE transport long the armchair (*xx*) and zigzag (*yy*) directions. Then, the LS is formed by replacing two S atoms with Se in a 2×1 supercell of the orthorhombic crystal, forming a superlattice period along *xx*, as shown in Figure 1c. Based on our previous study,[23] the $\kappa_l$ in $MoS_2$/$MoSe_2$ LS is greatly minimized as compared to the pristine monolayers. Thus, we expect to alter the TE performance of $WS_2$ or $WSe_2$ monolayers through the similar mechanism.

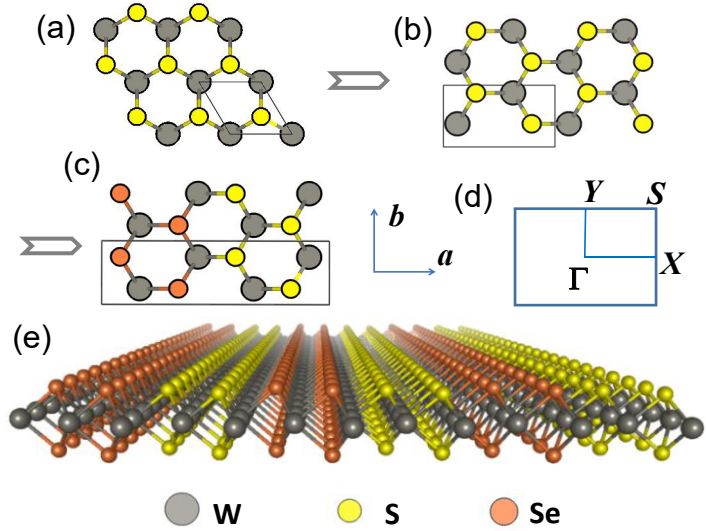

**Figure 1.** Schematic of the monolayer $WS_2$ and $WS_2$/$WSe_2$ LS. (a) and (b) show the schematic diagram of monolayer $WS_2$ with hexagonal crystal and rebuild orthorhombic crystal, respectively. (c) is the constructed $WS_2$/$WSe_2$ LS structure, which can be obtained by replacing two S atoms with Se in a 2×1 supercell of monolayer $WS_2$. Noting that the *a*-axis (*xx*) and *b*-axis (*yy*) are parallel (armchair) and perpendicular (zigzag) to the superlattice period, respectively. (d) shows the Brillouin zone high-symmetry points and (e) is the extended view of $WS_2$/$WSe_2$ LS.

The calculated SOC-MBJ band structure and DOS of monolayer $WS_2$ and $WS_2$/$WSe_2$ LS are shown in Figures 2(a) and 2(b), respectively. Calculated band gap 1.58eV for monolayer $WS_2$ is in agreement with reported value of 1.55eV[39] and it is the smaller 1.34eV for WS$_2$/WSe$_2$ LS. The uppermost valence band and the low-lying conduction bands relative to the Fermi level, which individually determine the hole and electron transport properties, both primarily arise from the W|$d$> and S|$p$> states for monolayer WS$_2$, while for WS$_2$/WSe$_2$ LS they in addition come form Se|$p$> states, as the projected DOS shown in Figure S1. A closer look over the band structure reveals that the SOC effect leads to markedly split of the conduction band while the valence band seems like unaffected, this SOC induced band splitting primarily arises from the $d$ orbital splitting of transition-metal W when it bonded to ligands S or Se, according to Li *et al*.[40] Importantly, the split leads to a three-fold degenerate lowest-energy $C_1$, $C_2$ and $C_3$ sub-bands in both monolayer WS$_2$ and WS$_2$/WSe$_2$ LS, as shown in Figure 2c, which leads to the high conduction DOS. According to Pei *et al*.,[41] a highly degenerate level of electronic band will optimize the electronic performance through the increased factor $N_v$, which in turn increases the DOS effective mass ($m^* = N_v^{2/3} m_b^*$) without explicity reducing carrier mobility $\mu$, as the $\mu$ is low for heavy band-mass. In particular, the degenerate three sub-bands separate in energy to the upper $C_4$ conduction band, giving rise to a remarkable stairlike DOS and playing an important role in electronic transport.

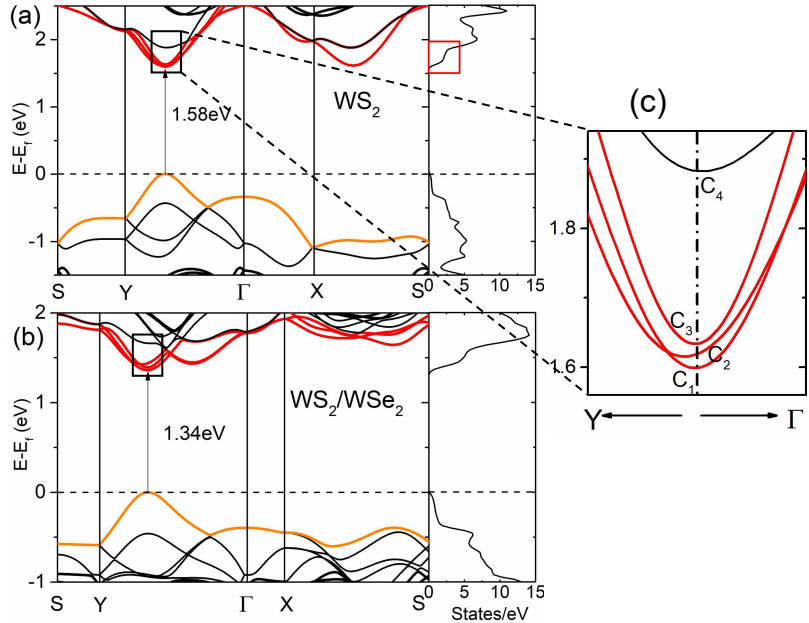

**Figure 2.** Calculated electronic band structure and density of states (DOS) of monolayer WS$_2$ (a) and WS$_2$/WSe$_2$ LS (b) along the symmetry directions of the Brillouin zone with SOC effect and MBJ functional included. (c) Conduction bands near the Fermi energy. The red box labeled in (a) shows the remarkable stairlike DOS at the conduction band minimum of monolayer WS$_2$.

The electronic transport coefficients are calculated by BTE within a CRTA. Within CRTA, the σ and $\kappa_e$ can only be calculated with a constant relaxation time $\tau$ included, while $\tau$ is canceled in Eq. (2) of $S$. Thus, $S$ can be directly calculated from the DFT band structure, but the evaluation of σ and $\kappa_e$ still require the knowledge of $\tau$.[42] Here, we employ the deformation potential (DP) theory[43] of Bardeen and Shockley to calculate $\tau$, which is generalized to include the scattering of carriers by acoustic modes and allows to evaluate τ in terms of the DP constant ($E_l$), elastic constant ($C_{2D}$) and effective mass ($m^*$). Based on these parameters, the carrier mobility can be obtained according to the formula:

$$\mu = \frac{e\hbar^3 C_{2D}}{k_B T m^* m_d^* E_l^2}, \qquad (1)$$

and the relaxation time τ is given by:

$$\tau = \frac{\mu m^*}{e}. \qquad (2)$$

Here, $C_{2D} = \frac{1}{S_0}\frac{\partial^2 E}{\partial(\Delta l/l_0)^2}$, a second order of the total energy with respect to deformation $\Delta l/l_0$, and $E_l = \frac{\partial E_{edge}}{\partial(\Delta l/l_0)}$, the slop of the band-edge energies with respect to $\Delta l/l_0$, and $m_d^* = \sqrt{m_{\Gamma-X}^* m_{\Gamma-Y}^*}$. Calculated results are shown in Figure S2 and Table 1.

Although the DP theory shown reliability for calculation of τ, we still need to discuss the possible challenges of this theory here since it just consider acoustic phonon scattering on carriers, which in some cases will overestimate τ. For WS$_2$/WSe$_2$ LS, the increase of atoms in the unit cell leads to a large number optical modes, as shown in Figure 4(b), and also the acoustic modes are coupled with low-lying optical modes. However, these optical modes are almost linearly dispersion except the ZO, TO, and LO modes, which means that possible scattering on carriers may arise from these low-frequency optical phonons. From monolayer WS$_2$ to WS$_2$/WSe$_2$ LS, the polar optical phonon (POP) scattering should be checked as the introduced small dipole. According to Hung et al.,[44] the strength of electron-POP coupling mainly arises from the phonons with frequency separation around LO and TO modes, which can be quantified by a ratio $\omega_L/\omega_T$, where $\omega_L$ and $\omega_T$ are the frequencies of LO and TO modes. A clear view of the LO and TO modes are shown in Figure S3. A significant frequency separation along Γ-X and Γ-Y is found for monolayer WS$_2$ while the separation along X-S-Y is significant for WS$_2$/WSe$_2$ LS. Taking an averaged $\omega_L$ and $\omega_T$ values, we determine the ratio $\omega_L/\omega_T$ for monolayer WS$_2$ and WS$_2$/WSe$_2$ LS are 1.09 and 1.08, respectively, which might explain the unconspicuous POP effect in LS crystal, as compared to monolayer WS$_2$. In addition, the τ usually decreases with the reduction of dimensionality due to the quantum confinement, in a non-polar case which could be reflected by the enhanced DOS near the Fermi level. Based on the Fermi golden rule for non-polar case, the relation $\tau^{-1} \propto$ DOS can be defined.[45] Therefore, the higher DOS near the Fermi level (see Figure 2(b)) indicates the enhanced quantum confinement in WS$_2$/WSe$_2$ LS, and hence leads to the decreased τ (see Table 1). Besides, the higher hole τ is in accordance with the lower valence DOS.

The calculated $E_l$ and $C_{2D}$, as shown in Table 1, are nearly independent of the structural orientation. Thus, the highly asymmetric carrier mobility along xx and yy comes from the anisotropic effective mass. It is regularly found that electrons with higher $m^*$ along yy leads to the lower μ in this direction, while holes instead with lower $m^*$ along yy leads to the higher μ in this direction. Although μ is highly anisotropic, the calculated relaxation time τ is almost isotropous since it only depends on $E_l$ and $C_{2D}$ with $m^*$ excluded according the Eqs. (4) and (5). Based on Mott relation, σ and S in Eqs. (1) and (2) can be simplified as[8]:

$$\sigma(\varepsilon) = n(\varepsilon)e\mu(\varepsilon) = n(\varepsilon)e^2 \frac{\tau(\varepsilon)}{m^*}, \quad (3)$$

$$S = \frac{\pi^2}{3}\frac{k_B^2 T}{q}\left(\frac{d[\ln(\sigma(\varepsilon))]}{dE}\right)_{E=E_F}. \quad (4)$$

For similar $\tau$, higher $\mu$ (lower $m^*$) means longer mean free path, which usually indicates a smaller carrier density $n(\varepsilon)$. Therefore, there could be a vanishing anisotropy in electronic transport. In addition, one can also notice from Table 1 that the DP results for monolayer WS$_2$ and WS$_2$/WSe$_2$ LS are much similar. For parabolic bands, the transport distribution can be defined in simpler form: $\sum(\varepsilon) = g(\varepsilon)v(\varepsilon)^2\tau(\varepsilon)$ [31], where $g(\varepsilon)$ is the DOS. As a result, the similar DOS, $\mu$, and $\tau$ could result in small anisotropic of electronic transport coefficients between monolayer WS$_2$ and WS$_2$/WSe$_2$ LS. Moreover, it is important to emphasize that the much lower $p$-type $E_l$, which reflects the minimal sensitivity of valence band maximum to deformation, leads to the extremely high $\mu$ for holes, as compared to electrons. It is expected to obtain a better $p$-type TE performance from this high hole mobility.

Electronic transport coefficients (300K) along $xx$ and $yy$ as a function of hole and electron density are shown Figure 3. We find here the nearly vanishing anisotropy and the much closer performance between monolayer WS$_2$ and WS$_2$/WSe$_2$ LS. It important to note that the $S$ shown in Figs. 3(a) and 3(b) are quite large at 300K with a value of 552 $\mu$V/K of monolayer WS$_2$ at a typical electron density $10^{19}$cm$^{-3}$. The $n$-type $S$ is higher than $p$-type primarily due to the high conduction DOS arising from the three-fold degenerate sub-bands, according to Pei et al.,[41] From Eqs. (6) and (7), the $S$ can be enhanced from an increased energy-dependence of carrier density $dn(\varepsilon)/d\varepsilon$, which can be achieved by enhancing the dependence of DOS on energy, $dg(\varepsilon)/d\varepsilon$.

Table 1. DP constant, elastic modulus, carrier effective mass, carrier mobility, and carrier relaxation time at 300 K. Both electrons and holes along different directions are calculated.

|  |  |  | $E_l$ (eV) | $C_{2D}$ (eVÅ$^{-2}$) | $m^*$ ($m_e$) | $\mu$ (cm$^2$V$^{-1}$s$^{-1}$) | $\tau$ ($10^{-14}$s) |
|---|---|---|---|---|---|---|---|
| WS$_2$ | n | xx | -9.25 | 9.15 | 0.27 | 229.37 | 3.51 |
|  |  | yy | -9.23 | 9.16 | 1.27 | 49.03 | 3.53 |
|  | p | xx | -2.44 | 9.15 | 1.01 | 799.87 | 45.78 |
|  |  | yy | -2.45 | 9.16 | 0.42 | 1909.91 | 45.46 |
| WS$_2$/WSe$_2$ | n | xx | -9.08 | 8.18 | 0.4 | 119.36 | 2.71 |
|  |  | yy | 8.18 | 8.31 | 1.27 | 40.57 | 2.92 |
|  | p | xx | -2.39 | 8.18 | 1.19 | 604.67 | 40.77 |
|  |  | yy | 2.47 | 8.31 | 0.39 | 1754.88 | 38.78 |

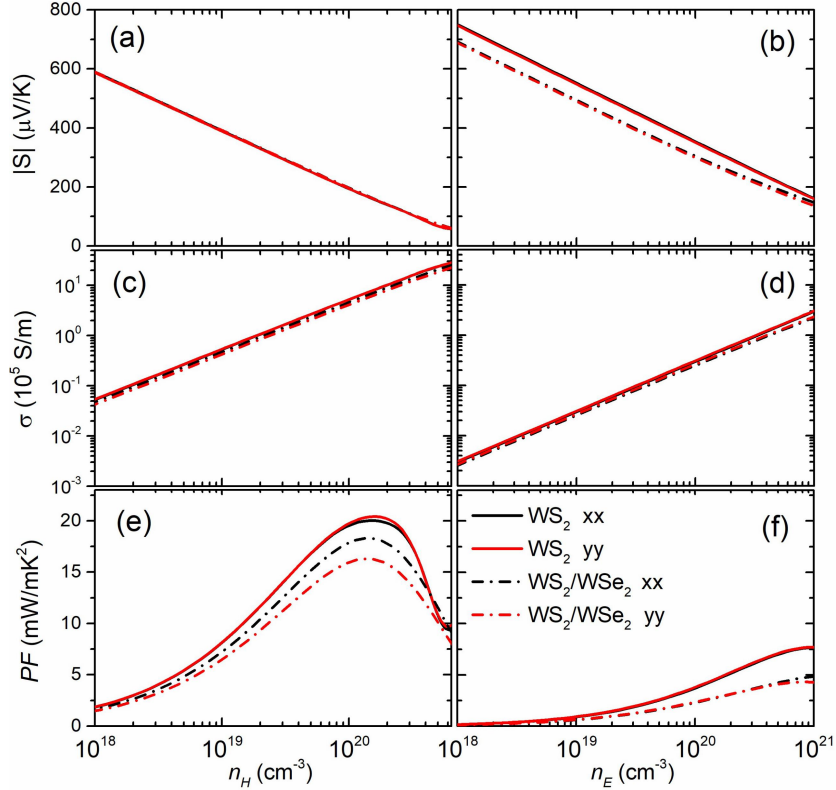

**Figure 3.** Electronic transport coefficients of monolayer WS$_2$ and WS$_2$/WSe$_2$ LS parallel (*xx*) and perpendicular (*yy*) to the superlattice period at 300K. The left and right panels denote the *p*- and *n*-doped performance, respectively. (a) and (b) the Seebeck coefficient (*S*), (c) and (d) the electrical conductivity (σ), and (e) and (f) the power factor (*PF*=$S^2$σ).

Consequently, the prominent stairlike conduction DOS of monolayer WS$_2$ accounts for its higher *n*-type *S*, as shown in Figure 3b. In contrast to *S*, the *p*-type σ is much higher than *n*-type due to the extremely high $\mu$ of holes, see in Figs. 3(c) and 3(d). It is also detectable that the σ of monolayer WS$_2$ is a little higher than WS$_2$/WSe$_2$ LS due to its higher $\mu$. Finally, the power factors (*PF*=$S^2$σ) are given in Figs. 3(e) and 3(f), it is interesting that high *n*-type *S* as originated from the three-degenerate sub-bands doesn't overcome the lower *n*-type σ and leads to the low *n*-type *PF*. Instead, due to the extremely high hole mobility, the high *p*-type σ accounts for the high *p*-type *PF*. The optimal *PFs* are about 20.5 mW/mK$^2$ and 18.4 mW/mK$^2$ for *p*-doped monolayer WS$_2$ and WS$_2$/WSe$_2$ LS, respectively, which are higher than monolayer MoS$_2$,[15] WSe$_2$,[17] and even the star bulk material Bi$_2$Te$_3$.[29] The calculated $\kappa_e$ shown in Figure S4 is less than 0.5 W/mK at a carrier density $10^{19}$ cm$^{-3}$, thus the thermal conductivity should be dominated by lattice contribution.

Now, we set out to discuss the phonon transport properties in monolayer WS$_2$ and WS$_2$/WSe$_2$ LS. Like other 2H TMDC monolayers, monolayer WS$_2$ and WSe$_2$ possess high $\kappa_l$ also because of the wide frequency gap and the strong bonding stiffness, which leads to the long relaxation time and high group velocity. The theoretical $\kappa_l$ of WS$_2$ and WSe$_2$ monolayers at 300K are higher than 100 W/mK and 50 W/mK,[17, 39] respectively. It is worthwhile to note that these values are based on the hexagonal crystals (Figure 1a) and a single-mode RTA of BTE. The smaller span of phonon frequency, as arising from the heavier atomic mass of Se and the smaller elastic constant, leads to the lower $\kappa_l$ of WSe$_2$ monolayer. In order to minimize $\kappa_l$, it is desirable to manipulate the frequency gap by alloying the two pristine monolayers, which could be realized by using LS structure as building block.

The calculated phonon spectrum for monolayer WS$_2$ and WS$_2$/WSe$_2$ LS are shown in Figure 4 a and b, respectively. Due to the formula (WS$_2$)$_2$ of the rebuild orthorhombic monolayer WS$_2$, there are three acoustic branches and fifteen optical branches, which is a little different from the one based on hexagonal crystal.[39] The lowest ZA acoustic branch corresponds to the out-of-plane displacement, while the higher TA and LA branches correspond to the in-plane transverse and longitudinal displacements. In general, the acoustic branches do most of the contribution to $\kappa_l$ while the large number of optical branches instead have little contribution. Based on Ouyang *et al*.,[39] the optical contribution to $\kappa_l$ in monolayer WS$_2$ is less than 1%. Due to the large atomic weight difference between W and S, there is a wide frequency gap between acoustic branches and high optical branches, which partly forbids the anharmonic processes. However, for WS$_2$/WSe$_2$ LS, one can see that the acoustic branches are coupled with low-lying optical branches.

The $\kappa_l$ of monolayer WS$_2$ along armchair (*xx*) and zigzag (*yy*) directions are 33.8 W/mK and 9.2 W/mK, respectively, as plotted in Figure 4c. The *xx* and *yy* are both the principle axis of the $\kappa_l$ tensor because of the orthorhombic symmetry crystal. The $\kappa_l$ is inversely proportional to temperature $\kappa_l \propto 1/T$ since the enhanced phonon-phonon interaction with the increasing temperature. In contrast to the isotropic $\kappa_l$ in hexagonal crystal, we obtain here the strong anisotropic $\kappa_l$ along *xx* and *yy*, which can be better understood from the anisotropy of group velocity (*v*). The *v* of ZA, TA, and LA acoustic branches are plotted in Figure 4e, it is clear that the *v* has significant difference along *xx* and *yy* directions for all three acoustic branches. Both ZA and TA branches exhibit higher *v* along *xx* direction, while the TA branch shows higher *v* along *yy* direction. At low-frequency limit, the *v* of quadratic ZA branch approaches to zero at Γ point, and the *v* in *xx* and *yy* directions are 36 and 41 (×$10^2$ m/s) for TA branch, and 56 and 52 (×$10^2$ m/s) for LA branch, with an anisotropy of 1.1 and 1.08, respectively. The *v* of LA branch is quite higher than ZA and TA branches, which plays dominate contribution to $\kappa_l$. Therefore, the anisotropy of $\kappa_l$ arises from all three acoustic branches and the higher $\kappa_l$ in *xx* direction is mainly contributed by the ZA and LA branches.

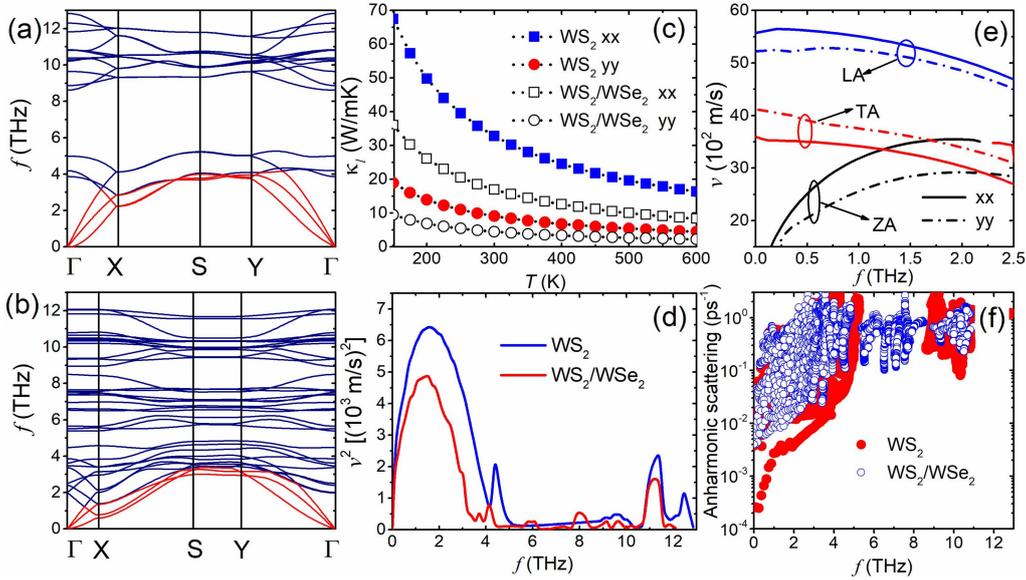

**Figure 4.** Phonon transport coefficients and performance. (a) and (b) are the calculated phonon dispersion of monolayer WS$_2$ and WS$_2$/WSe$_2$ LS, the two transverse (ZA, TA) and one longitudinal (LA) acoustic branches are highlighted in red. Calculated lattice thermal conductivity with respect to temperature is given in (c), and (d) shows the averaged square of phonon velocity of monolayer WS$_2$ and WS$_2$/WSe$_2$ lateral superlattice. (e) Acoustic phonon velocity (ZA, TA, LA) of monolayer WS$_2$ along different directions. (f) Comparison of anharmonic phonon scattering rate at 300K.

The calculated $\kappa_l$ for $WS_2$/$WSe_2$ LS at 300K, as presented in Figure 4c, are 16.9 W/mK and 4.5 W/mK along *xx* and *yy* directions, respectively. It is important that these values are nearly halved as compared to monolayer $WS_2$, which can be ascribed to the enhanced anharmonic three-phonon processes coming from the enhanced coupling between acoustic and optical branches. From Figure 4f, one can notice that the scattering probability of acoustic phonons and low-lying optical phonons (below 5 THz) in $WS_2$/$WSe_2$ LS are remarkably higher than that in monolayer $WS_2$. The enhanced anharmonic processes usually include the absorption and emission processes, in which two incident phonons with energy combined into one phonon, or one incident phonon is split among two phonons. In addition, we calculate the averaged square of *v*, $v_\omega^{2,\alpha} = \sum_\lambda v_\lambda^{2,\alpha} \delta(\omega-\omega_\lambda) / \sum_\lambda \delta(\omega-\omega_\lambda)$,[46] as plotted in Figure 4d. Obviously, the $v_\omega^2$ of $WS_2$/$WSe_2$ LS is quite lower than monolayer $WS_2$, which can be understood by the weaker bonding stiffness in $WS_2$/$WSe_2$ LS as confirmed by its smaller $C_{2D}$ in table 1. A weaker bonding stiffness causes a smaller span of frequency and leads to the smaller group velocity. It can be also detected from Figure 4d that the low-frequency phonons (below 5 THz) dominate the contribution to $\kappa_l$. As a result, we achieve the minimized $\kappa_l$ in $WS_2$/$WSe_2$ LS as compared to the pristine monolayers.

In order to discuss the anisotropic $\kappa_l$ in terms of the chemical bonding strength, we calculate the longitude IFCs of corresponding bonds in the unit cell, as labeled in Figure S5. In the finite displacement method, potential energy is represented as a function of atomic position $V[r(j_1),...,r(j_n)]$, based on which the IFCs is given by

$$\Phi_{\alpha\beta}(j,j') = \frac{\partial^2 V}{\partial r_\alpha(j) \partial r_\beta(j')}, \quad (5)$$

where *α* and *β* are the Cartesian indices, *j* and *j'* are the indices of atoms, and $\partial r_\alpha(j)$ is the finite displacement of *j*-th atom along *α* direction. Calculated IFCs for both monolayer $WS_2$ and WS$_2$/WSe$_2$ LS are shown in Table 2. $\Phi_1$ corresponds to IFC of bond 1 along *xx* (armchair) direction, while $\Phi_2$ and $\Phi_3$ correspond to IFC of bond 2 and 3 along *yy* (zigzag) direction. A large IFC means a strong chemical bonding stiffness, which leads to the severe atomic vibration and high $\kappa_l$. As shown in Table 2, in comparison to $\Phi_2$ and $\Phi_3$, the higher $\Phi_1$ is in accordance with the higher $\kappa_l$ along *xx*. While these IFCs are weakened when forming a WS$_2$/WSe$_2$ LS, which also points to the enhanced anharmonicity and minimized $\kappa_l$ in LS crystal.

Table 2. The interatomic force constants (IFCs) for monolayer WS$_2$ and WS$_2$/WSe$_2$ LS. $\Phi_1$, $\Phi_2$, and $\Phi_3$ are the IFCs for bond 1 along *xx*, bond 2 and bond 3 along *yy*, respectively.

|  | $\Phi_1$ (Ry/bohr$^2$) | $\Phi_2$ (Ry/bohr$^2$) | $\Phi_3$ (Ry/bohr$^2$) |
|---|---|---|---|
| WS$_2$ | 5.461 | 4.685 | 4.685 |
| WS$_2$/WSe$_2$ | 4.812 | 4.038 | 4.408 |

With all TE coefficients available, we now evaluate the figure of merit *zT* for both monolayer WS$_2$ and WS$_2$/WSe$_2$ LS. Figure 5 shows the evaluated *zT* at 300K as a function of carrier density. It is found that that *p*-type *zT* is much higher than *n*-type due to the excellent hole transport performance. Importantly, we achieve the enhanced *zT* in WS$_2$/WSe$_2$ LS as contributed by the minimized $\kappa_l$. The lower $\kappa_l$ in *yy* direction leads to the better TE performance. The optimal *p*-type *zT* of monolayer WS$_2$, as shown in Figure 5a, are 0.16 and 0.47 along *xx* and *yy*, respectively, which are higher than that of monolayer MoS$_2$,[15] MoSe$_2$ and WSe$_2$[17] (less than 0.1). In WS$_2$/WSe$_2$ LS, the optimal *p*-type *zT* increase to 0.28 and 0.68 along *xx* and *yy*, respectively, this value (0.68) is comparable to the maximum *zT* (~0.7) of commercial PbTe,[47] which is also a record value among 2H TMDCs monolayers. The inset in Figure 5b shows the optimal *p*-type *zT* of WS$_2$/WSe$_2$ LS with respect to temperature. With the increasing temperature, the optimal *zT* along *xx* and *yy* can exceed 1 at a threshold temperature around 550K and 400K, respectively, indicating the potential as candidate for *p*-type TE material working under moderate temperature. It is known that the experimental thermal conductivity is usually lower than theoretical one since the experimental samples inevitably contain defects, distortion and impurities. Despite the possible overestimation of $\tau$ by DP theory, we expect that such promising *zT* could be realized experimentally.

Before closing, the following three facts should be noted: (i) It is believed that the band degeneracy is an important feature for enhanced *PF*, as the DOS effective mass is enhanced through a factor $N_v$ which doesn't reduce $\mu$, based on this principle the *S* is enhanced without overshadowing the σ.[41,51] Herein, we show that this theory is eclipsed in low deformation potential (DP) crystal in which the carriers are less scattered by acoustic phonons and thus obtain high mobility. As we find in monolayer WS$_2$ and WS$_2$/WSe$_2$ LS, in spite of the highly degenerate conduction band, the low *p*-type DP leads to much high hole mobility and therefore the high *p*-type *PF*. (ii) Conventional bulk TE materials such as $Bi_2Te_3$ and PbTe point to that the features of high TE performance should be narrow band-gap (high PF) and heavy elements (low $\kappa_l$),[48-50] our calculations show that 2D crystals with large band-gap and containing light element also possess comparable *zT*, which demonstrates that low-dimension can modify the contradiction of TE coefficients. (iii) Finally, it is necessary to note the limitation of our model because the TE transport coefficients especially the $\kappa_l$ depend strongly on periodic length of the LS. However, relative reports on in-plane LS structure are rare. According to early reports on bulk $Bi_2Te_3/Sb_2Te_3$,[52] Si/Ge superlattice,[53] and even the Si/Ge superlattice nanowires,[54] the variation of $\kappa_l$ shows first reduction and latter growth as the increase of periodic length. This bipolar behavior can be understood from two mechanisms, that is, the lattice mismatch introducing anharmonic effect at the interface and depressing the phonon velocities and $\kappa_l$, the increase of periodic length however gradually leading to a phonon localization with the reduced scattering. Based on Broido and Reinecke,[53] this process can defined as a competition between the flattened phonon dispersion that inhibits heat conduction and the reduced umklapp scattering that instead enhances it, as the periodic length rises from the ultrathin limit to a larger scale. Therefore, by optimizing the periodic length, one can expect to obtain the best TE performance of superlattice based bulks or monolayers. In addition, Janus monolayers and their LS were also proposed,[55,56] much lower $\kappa_l$ (below 15 W/mK for SMoSe at 300K)[57] was found as compared to monolayer TMDCs, hence, it is desirable to achieve high TE performance by alloying Janus monolayes with LS.

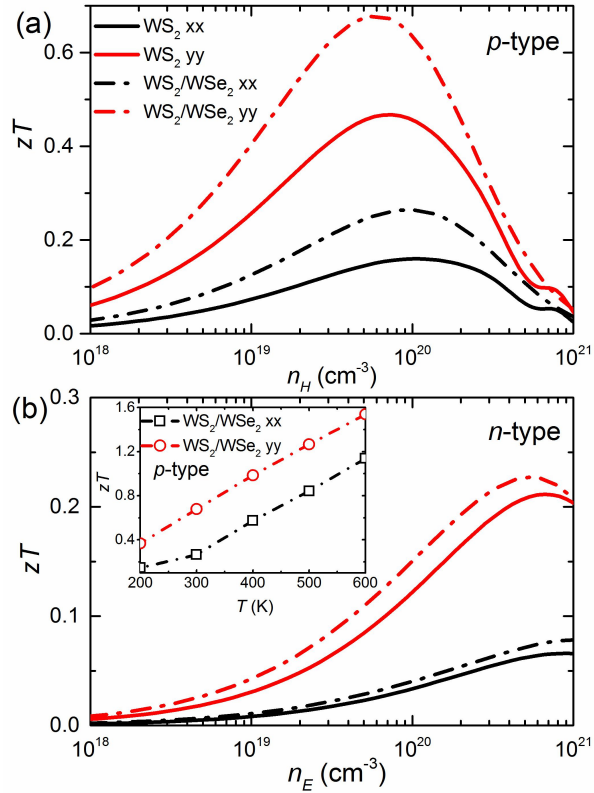

**Figure 5.** The calculated thermoelectric figure of merit $zT$ at 300K as a function of hole (a) and electron (b) density. The inset in (b) shows the temperature-dependent optimal $p$-type $zT$ of $WS_2$/$WSe_2$ LS.

## CONCLUSION

In summary, we propose to improve the TE performance of 2H TMDCs monolayers using LS crystal as the configuration strategy based on the experimental report. The TE transport properties of monolayer $WS_2$ and $WS_2$/$WSe_2$ LS are investigated by DFT combined with BTE. The underlying origins of electronic transport lie in the electronic band structure consisting of (i) the three-fold degenerate sub-bands contribute to the high conduction DOS and therefore the high $n$-type Seebeck coefficient, (ii) the remarkable stairlike DOS relative to the Fermi level accounts for the higher $n$-type Seebeck coefficient of monolayer $WS_2$, (iii) the high $p$-type σ arising from the extremely high hole mobility leads to the high $p$-type power factor. Besides, the symmetric electronic transport along zigzag ($xx$) and armchair ($yy$) directions can be attributed to the nearly isotropic lifetime of charge carriers. Minimized lattice thermal conductivity in $WS_2$/$WSe_2$ LS can be understood from the enhanced anharmonic processes and also the weaker bonding stiffness. The phonon anisotropy is explained from the phonon dispersion and group velocity. Due to the remained electronic transport and the minimized lattice thermal conductivity, we achieve the enhanced $zT$ in $WS_2$/$WSe_2$ LS with the optimal $p$-type value exceeds 1 around 400K. Our results pave the way towards opportunities for enhancing the TE performance of extensive 2D materials by using LS as building block.

## AUTHOR INFORMATION

Corresponding authors
*Email: dinggq@cqupt.edu.cn (Guangqian Ding)
*Email: Cumt.dun@gmail.com (Chaochao Dun)
*Email: wangxt45@126.com (Xiaotian Wang)
**Notes**
There are no conflicts to declare.

## ACKNOWLEDGMENT

This work is supported by National Natural Science Foundation of China (Grant No.11804040) and Fundamental Research Funds for the Central Universities (Grant CQUPT: A2017-119).